# Improved three component IVIM fitting using physics-informed neural network with multimodal spatial prior

[a]Nhat Hoang, [b]Abrar Faiyaz, [c]Emmanuel A. Mensah, [b,c,d] Md Nasir Uddin, [a,c,d]Jianhui Zhong, [b,d,e]Giovanni Schifitto

[a]Department of Physics, University of Rochester, Rochester, NY 14627, USA
[b]Department of Neurology, University of Rochester, Rochester, NY 14642, USA
[c]Department of Biomedical Engineering, University of Rochester, Rochester, NY 14627, USA
[d]Department of Imaging Sciences, University of Rochester, Rochester, NY 14642, USA
[e]Department of Electrical & Computer Engineering, University of Rochester, Rochester, NY 14627, USA

Corresponding author: Nhat Hoang (nhat_hoang@urmc.rochester.edu)

Present address: 601 Elmwood Avenue, Box 67, Rochester, NY 14642

**Working Title:** Specialized multimodal physics-informed network for three-component IVIM fitting

**Abstract Word Count**: 233

**Manuscript Word Count**: 4850

**Number of Primary Tables and Figures**: 10

**Number of Supplementary Figures and Tables**: 5

**Total Page Count**: 30

## Abstract

**Purpose**: To mitigate parameter degeneracy and improve robustness of IVIM fitting, a multimodal physics-informed neural network was developed to leverage anatomical information through the combined use of cross attention, selective spatial gating, and anatomy-weighted spatial prior.

**Methods**: T1w and FLAIR anatomical images were co-registered to IVIM signal in native space and segmented by tissue type. The proposed Physics-Anchored Cross-attention Embedding network (PACE) encoded IVIM signals and their anatomical features in parallel branches, then integrated them via cross attention and spatial gating. An anatomy-weighted spatial prior further constrained signal reconstruction using anatomically similar neighboring voxels. PACE and its prior-guided variant were compared with two baseline deep neural networks (DNN and IVIM3C), least squares, non-negative least squares, and maximum a posteriori fitting models. All models were evaluated on signal and parameter reconstruction accuracy and lesion contrast fidelity using data generated from subjects drawn from the University of Rochester COVID cohort.

**Results**: PACE variants demonstrated lower RMSE for all IVIM parameters except for $D_{mv}$. The anatomy-weighted spatial prior further reduced RMSE across all five parameters relative to the default PACE, with modest reductions in lesion contrast accuracy for $f_{mv}$ and $f_{int}$. For in-vivo evaluation, PACE with spatial prior also showed the best signal reconstruction across all noise levels while maintaining good to excellent ICC.

**Conclusion**: Multimodal anatomical information improved IVIM fitting accuracy and robustness, with selective spatial integration via cross attention, gating, and anatomy-weighted spatial prior.

## Introduction

Intravoxel incoherent motion (IVIM) refers to the microscopic movement of water molecules within a voxel during diffusion MRI acquisition. By characterizing signal attenuation over a range of diffusion-sensitive gradients *b*, IVIM can provide quantitative information about the microstructure and microvascular perfusion across different tissue types[1]. Since its introduction, IVIM has become a valuable imaging tool in oncology[1-8], multiple sclerosis[9], stroke[10, 11], inflammation[7, 12], and other neurological conditions[10, 13, 14], due to its sensitivity to changes in diffusion properties in many diseased processes.

The traditional bi-exponential IVIM model is composed of two components, one representing the diffusion of free water through tissues, and the other representing the perfusion of blood through capillary beds[1]. Recently, there has been a growing effort to expand the traditional IVIM model to three compartments to analyze complex tissue environments in the brain[14-18], where an intermediate diffusion component between the diffusivity ranges of parenchymal free water and microvascular perfusion was attributed to interstitial fluid observed in lesion-prone regions[16]. Subsequent studies have linked this intermediate component to impaired glymphatic clearance[12, 13], blood brain barrier (BBB) leakage[14], and cerebral small vascular disease (CSVD) [10, 14, 18, 19].

However, estimating IVIM parameters from diffusion data remains a formidable challenge, as it involves solving an ill-posed inverse problem. Because a single attenuation equation must resolve multiple parameters simultaneously, different combinations of IVIM parameters can produce near identical signal profiles. This issue is compounded by noise presence in the in-vivo data, leading to instability in perfusion parameters[5]. Traditional methods such as least squares (LSQ) are particularly prone to noise corruption[4, 20-22]. For three-component IVIM, this issue is further exacerbated by the additional parameters, raising the need for better fitting precision in low signal-to-noise (SNR) regimes.

Currently, machine learning (ML) approaches such as deep neural net (DNN) models have proven to be an attractive alternative, demonstrating improvements in both computational scalability and noise resilience over traditional methods for both two and three component IVIMs[4, 18, 22-24]. However, ML performance is influenced by their architecture designs and training strategies: unsupervised DNNs can overfit to noise, while supervised DNNs are biased toward the training population[21].

In recent years, multi-modal approaches in diffusion MRI have successfully leveraged complementary information to enhance image reconstruction robustness and diagnostic accuracy compared to unimodal methods[25-27], yet it remains unknown how such context should be integrated to improve IVIM fitting. Fundamentally, imaging contrast arises from the interaction between tissue's physical properties and the specific imaging modality. Consequently, different tissue subtypes are likely to exhibit sufficiently distinct combination of T1-weighted (T1w), Fluid-Attenuated Inversion Recovery (FLAIR) profiles to break the parameter degeneracy for similar IVIM attenuations. To explore this idea, we construct **P**hysics-**A**nchored **C**ross-attention **E**mbedding net (PACE) to leverage spatial information and improve IVIM reconstruction accuracy and tissue contrast fidelity. Specifically, we integrate PACE's local anatomical context with IVIM signal through convolutional network encoding, while using an anatomy-weighted spatial prior to regularize the signal reconstruction loss.

# Methods

### 2.1 Study Population

A total of 120 subjects were selected from an observational study assessing neurological sequelae of long COVID to construct our synthetic and in-vivo training and testing data. The cohort (mean age 69.3 ± 5.1 years, 69 females, 51 males) included 100 individuals with varying white matter hyperintensities (WMH) severity based on Fazekas scores (36 grade 1, 19 grade 2, 45 grade 3), with WMH volume median of 684 $mm^3$ and WMH interquartile range 194 to 2328 $mm^3$. Seventy-five subjects had a history of COVID-19 (57 mild, 18 severe). The study was approved by the University of Rochester's Research Subject Review Board, with all enrolled subjects provided consent before data were taken.

### 2.2 MRI acquisition

MRI images were acquired on a Siemens 3T Prisma Fit scanner (Germany) using a 64-channel head coil. T1w images were obtained with inversion time (TI) = 926 ms, repetition time (TR) = 1840 ms, echo time (TE) = 2.45 ms, and a 1 mm isotropic resolution. FLAIR images were acquired with TI = 1800 ms, TR = 5000 ms, TE = 100 ms, and a 1 mm isotropic resolution. IVIM diffusion-weighted imaging used a spin-echo EPI sequence (TR/TE = 6800/84 ms) with 2.4 mm isotropic resolution, with GRAPPA in-plane acceleration factor of 2 and fifteen diffusion b-values ($b$ = [0, 5, 7, 10, 15, 20, 30, 40, 50, 60, 100, 200, 400, 700, 1000] s/mm²) applied in 3 orthogonal directions.

### 2.3 IVIM models

The three-component IVIM model describing the normalized signal decay curve is defined as follows:

$$S_{norm} = \frac{S(b)}{S(0)} = f_{par}e^{-bD_{par}} + f_{mv}e^{-bD_{mv}} + f_{int}e^{-bD_{int}} \quad (1)$$

where $S_{norm}$ is the normalized IVIM signal, *S(b)* is the voxel signal measured at b-value *b*. $D_{par}$, $D_{int}$ , and $D_{mv}$ are the parenchymal, intermediate, and microvascular diffusivities measured in $mm^2/s$, respectively. The coefficients $f_{int}$ and $f_{mv}$ are the signal fractions associated with the intermediate and microvascular components. The parenchymal fraction $f_{\text{par}} = (1 - f_{mv} - f_{int})$ is constrained by the other two fractions, and therefore not expressed explicitly following previous convention[18].

### 2.4 Training Data Processing

#### 2.4.1 Synthetic patch creation

Fifty-two out of 120 subjects were reserved for synthetic data generation. To estimate the IVIM parameter ground truth, approximately 2.2 million voxels selected from 6 subjects with the highest WMH lesion burden were fitted via LSQ and grouped by tissue class GM, NAWM, and WMH using tissue and lesion maps produced by LesionBrain[28, 29]. These fits established the tissue-specific parameter ranges (Table 1) from which synthetic parameters were sampled using log-normal distribution. A fifth of the patches were planted as cross-tissue degenerate pairs whose IVIM attenuation signals differ by less than 0.2% in signal root mean square (RMS) across b-values.

For each synthetic voxel, 3x3 and 5x5 neighborhoods of T1, FLAIR, b0 and normalized IVIM neighborhood signal (IVIM_nb) were extracted from a tissue-matched location of one of the 52 reserved subjects to provide the spatial context for the center voxels.

The final training data set consists of 570k voxels (400k GM, 160k NAWM, 10k WMH), with validation and test sets both consist of 38k voxels (25k GM, 10k NAWM,3k WMH). All voxels in the synthetic data set are unique, with no overlap across data splits.

#### 2.4.2 In-vivo patch creation

For in-vivo inference, tissue-labeled T1w and FLAIR were simply co-registered to native IVIM space to generate the spatial features. For each target voxels, a 3×3 neighborhood of T1, FLAIR, b0, and IVIM_nb patch was extracted to provide spatial context for IVIM fitting.

### 2.5 Neural Network Architecture

Our network consists of a primary signal encoder for the center IVIM signal and separate spatial CNN encoders for the T1w and FLAIR neighborhood patches. The features of each encoder are then integrated through a cross-attention block to provide spatial context for IVIM fitting. The resulting spatial information is then modulated by the gating mechanism for each individual IVIM parameter, allowing the contribution of spatial context to be selected individually for each IVIM parameter. The network is then optimized using IVIM signal reconstruction and tissue classification losses. To resolve signal degeneracy, we also introduced an anatomy-weighted spatial prior implemented as a neighborhood signal consistency loss (section 2.6). The standard PACE network architecture overview is captured by Figure 1.

### 2.5.1 Modality representations

The IVIM encoder takes in the normalized IVIM signal from the target voxel and maps it to a latent vector $z_{ivim}$ to represent signal attenuation characteristics. The encoder consists of two fully connected layers with ELU activation, followed by a final linear projection and layer normalization.

The anatomical encoder consists of two architecturally identical CNN encoders dedicated to the T1w and FLAIR neighborhood patches. Each patch is processed through a two-layer CNN block with 3x3 kernels, stride 1, zero padding 1, and ELU activations. Their outputs features are flattened, linearly projected, and layer-normalized to produce respective latent presentations $z_{T1w}$ and $z_{FLAIR}$, which serve as spatial tokens for the cross-attention module.

### 2.5.2 Cross-attention and spatial context integration

The purpose of the cross-attention mechanism is to integrate multimodal spatial context given an anatomical neighborhood contrast and its corresponding IVIM signal. The query Q is constructed from the gradient-isolated copy of IVIM signal representation $z_{ivim,q}$ and a learned bias token τ, while the key K and value V matrices are identically constructed from spatial latent vectors $[z_{T1w}, z_{FLAIR}]$.

The attention score is represented by the multi-head cross-attention[30]

$$\mathrm{CrossAttn}(Q, \mathrm{K}, \mathrm{V})_{\mathrm{h}} = \mathrm{softmax}\left(\frac{\mathrm{Q_h K_h^T}}{\sqrt{\mathrm{d_k}}}\right)\mathrm{V_h}, \tag{2}$$

where $d_k = 16$ is the embedding dimension for each attention head $h = \{1,2\}$. The output of each head is then concatenated and linearly projected to obtain the total cross-attention output. The spatial contribution term $\delta_{spatial}$ is thus defined as

$$\delta_{spatial} = \mathrm{LayerNorm}\big(\alpha \cdot \mathrm{CrossAttn}(Q, K, V)\big), \tag{3}$$

where α is a learned scaling factor. The gradient-isolated $z_{ivim,q}$ in Q prevents spatial information from updating the IVIM encoder. Instead, IVIM parameters are selectively modulated by $\delta_{spatial}$ through the gating mechanism described below.

### 2.5.3 IVIM parameter gating mechanism

The gating mechanism regulates the contribution of spatial information from attention layer to each IVIM parameter head via individualized gating. For each IVIM parameter, their latent representation $z_X$ is computed as

$$z_X = z_{ivim} + \sigma(g_X) \cdot \delta_{spatial}, \tag{4}$$

where $\delta_{spatial}$ is the spatial contribution from the attention layer, $\sigma(g_X)$is the sigmoid transformed gate controlling the amount of spatial information added, and $g_X$ is its tuning constant. The schematic of this gating mechanism is illustrated by Figure S1.

During training, $\sigma(g_X)$ is initialized to $\sigma(-2)$ for all parameters. By construction, the gates only receive spatial context when it improves the signal fit, preventing fusion output from overwhelming the primary physics-anchored signal. To preserve biological plausibility, unity constraint $f_{par} + f_{int} + f_{mv} = 1$ was applied at the end of spatial gating.

### 2.6 Anatomy-weighted spatial prior

To explicitly test whether anatomical information can help distinguish IVIM signal degeneracy, we introduced an anatomy-weighted prior that constrains the predicted center-voxel signal using anatomically similar neighbors. For center voxel $i^{th}$ and its $j^{\text{th}}$ neighbor, anatomical distance is defined as the difference in their anatomical signal intensities:

$$d_{ij} = \sqrt{\left(\text{T1w}_i - \text{T1w}_j\right)^2 + \left(\text{FLAIR}_i - \text{FLAIR}_j\right)^2}. \tag{5}$$

The weight of the signal is then calculated according to their anatomical similarity:

$$w_{ij} = \frac{\exp\left(-d_{ij}^2/h^2\right)}{\sum_{k\in\mathcal{N}(i)} \exp\left(-d_{ik}^2/h^2\right)}, \tag{6}$$

where $h = 0.5$ is chosen to modulate the weight decay due to anatomical dissimilarity. The anatomical-weighted neighborhood signal is then given as

$$\bar{S}_i(b) = \sum_{j\in\mathcal{N}(i)} w_{ij} \frac{b_{0,j}}{b_{0,i}} S_j(b), \tag{7}$$

where $S_j(b)$ is the normalized neighbor IVIM signal and $\frac{b_{0,j}}{b_{0,i}}$ is ratio of between the neighbor and center voxel raw signal amplitudes.

The rationale behind (7) is that constraining the predicted signal with surrounding anatomical context can help resolve signal degeneracy without directly imposing constraints on the IVIM parameters. The spatial prior objective loss is thus

$$\mathcal{L}_{\text{spatial}} = \text{MSE}\left(\hat{S}_{pred}(b), \bar{S}_i(b)\right). \tag{8}$$

The procedure of selected spatial prior is illustrated in Figure 2.

### 2.7 Training Objective

The IVIM loss $\mathcal{L}_{\text{IVIM}}$ is defined as the mean square error (MSE) between the predicted IVIM signal and the normalized IVIM ground truth signal. Beside $\mathcal{L}_{\text{IVIM}}$, the network also incorporates the auxiliary losses $\mathcal{L}_{CE}$ and $\mathcal{L}_{spatial}$ (defined above) for regularization. $\mathcal{L}_{CE}$ is the weighted cross-entropy loss for network's tissue classification errors and was included to encourage the spatial encoders to learn tissue-discriminative features. A separate classification head, distinct from the attention block, predicts tissue type from spatial features with neither tissue labels nor classifier output directly entering the cross-attention or spatial pathway.

The total loss function is

$$\mathcal{L}_{total} = \mathcal{L}_{IVIM} + \alpha_{spatial}\mathcal{L}_{spatial} + \alpha_{CE}\mathcal{L}_{CE}, \quad (9)$$

where $\alpha_{spatial}$ and $\alpha_{\mathrm{CE}}$ are tuning strengths of the auxiliary losses. The default PACE setting used $\alpha_{CE} = 0.2$. and $\alpha_{spatial} = 0.1$ when spatial prior was enabled, so that the auxiliary losses are approximately an order of magnitude lower than $\mathcal{L}_{IVIM}$.

## 2.8 Model Training Procedure

### 2.8.1 Synthetic Training

Three deep learning models (PACE, DNN, and IVIM-3C[18]) were trained on the synthetic data (section 2.4.1) at multiple SNRs (20, 25, 30, and 35 dB) using the hyperparameter setting summarized in Table 2. To simulate the SNRs, Rician noise was injected into the synthetic IVIM signal of each targeted voxel by adding zero-mean Gaussian noise to the signal's real and imaginary components, with $\sigma = 10^{-\mathrm{SNR}/20}$, followed by magnitude reconstruction. Since in-vivo T1w/FLAIR inputs already contained noise from acquisition, no additional noise were added.

PACE and signal-only DNN are architectural variations of the same network and therefore use identical setting apart from the spatial configuration, while IVIM-3C[18] is kept its default setting as an external published method (Table 2). When spatial setting is enabled, tissue labels enter PACE through $\mathcal{L}_{CE}$ only during training, otherwise no labels were used at inference. When enabled, spatial prior acts as an additional signal regularization loss on neighborhood anatomical similarity, and likewise only active during training.

All networks used the same parameter bounds defined in Table 2, with the sigmoid range widened by 30% of each span to reduce saturation near the boundaries.

### 2.8.2 In-vivo evaluation

For vivo evaluation, networks trained on synthetic data at 25 SNR were applied directly to the independent test in-vivo data without retraining or fine-tuning to evaluate generalizability of the synthetically trained nets. IVIM signals were normalized by the median $b_0$ signal, while T1w and FLAIR intensities were clipped to their $1^{st}$-$99^{th}$ percentile range and scale to [0, 1] to be consistent with the network preprocessing.

### 2.8.3 Conventional Models

Segmented LSQ fitting and NNLS were conducted as described in previous publications[15, 18].

LSQ used a two-step approach, first estimating $D_{par}$ from signals with b ≥ 200 s/mm$^2$, followed by bounded nonlinear fitting using *SciPy*'s curve fit. NNLS used a 200-point logarithmically spaced diffusion spectrum, with spectral components grouped according to predefined diffusivity ranges. MAP optimization used empirical priors derived from the initial LSQ estimates followed by *L-BFGS-B* optimization as described in previous works[2, 4, 22]. All conventional methods were fitted with the same parameter bounds as described in section 2.8.1.

## 2.9 Evaluation Metrics

IVIM fitting models were evaluated based on three categories: reconstruction accuracy, lesion contrast fidelity, and in-vivo inference stability.

### 2.9.1 Reconstruction Accuracy

Parameter estimation accuracy was measured as the RMSE between ground-truth IVIM parameters and the predicted IVIM parameters for synthetic data. Signal fidelity was calculated using the RMSE between the ground-truth signal and the reconstructed IVIM signal for both in-vivo and synthetic tests.

### 2.9.2 Lesion Contrast Fidelity

Clinical WMH-NAWM conspicuity was quantified using the CNR contrast between WMH and NAWM for each of the IVIM parameters, given by the lesion WMH-NAWM contrast equation defined below[31]:

$$\text{Lesion CNR}_X = \frac{\left|\mu_X^{\text{WMH}} - \mu_X^{\text{NAWM}}\right|}{\sqrt{\left(\sigma_X^{\text{WMH}}\right)^2 + \left(\sigma_X^{\text{NAWM}}\right)^2}}\ , \tag{10}$$

where $\mu_X$ and $\sigma_X$ denote the mean and standard deviation of the IVIM parameter *X* within the respective tissue class.

### 2.9.3 Intraclass Correlation Coefficient

In-vivo inference stability was assessed using the two-way mixed, single measurement ICC formula[32] computed with five different seed initializations at different noise levels across all 38 in-vivo test subjects. The higher ICC rating indicates better inference stability, with values below 0.5 indicating poor, 0.5– 0.75 moderate, 0.75–0.90 good, and above 0.90 excellent.

## 2.10 Ablation Study

To isolate the individual contributions of additional spatial inputs, attention fusion, gating and auxiliary losses, the default PACE described in section 2.5 was ablated one component at a time (Figure 3, left configuration chart). The ablation arms were compared using parameter RMSE. Lesion CNR was also evaluated but only served as secondary criterion when param RMSE performance was comparable.

### 2.11 Model Comparison

#### 2. 11.1 Synthetic Comparison

To evaluate parameter estimation accuracy, signal IVIM parameter RMSE was computed for seven candidate models: PACE, PACE with spatial prior, DNN, IVIM-3C, LSQ, NNLS, and MAP at four Rician noise levels (20, 25, 30, and 35 dB, referenced at $b = 0$). To assess contrast fidelity, IVIM lesion CNRs and their corresponding lesion CNR RMSE were computed. To visualize the estimation bias, predicted WMH distribution was plotted against the ground truth distribution.

#### 2.11.2 Hybrid brain and WMH burden

To bridge the gap between in-vivo and synthetic data, a hybrid brain template was constructed from an in-vivo subject with the greatest WMH burden. We retained the subject's T1w, FLAIR, and $b_0$ data while assigned synthetic IVIM parameters to their voxels drawn from Table 1's LSQ distributions. Predicted parameter maps were evaluated at SNR 25 dB and compared against the IVIM synthetic ground truth.

#### 2.11.3 In-vivo Comparison

Because the IVIM parameter ground truth was unavailable in-vivo, reconstruction fidelity was evaluated using signal RMSE against the measured signal. This was computed over a stratified sample of 5000 voxels per test subject. For each subject, all available WMH voxels were included, and the remaining voxel allocations were divided equally between GM and NAWM. The different levels of Rician noise were added to each test subject's IVIM signal as described in 2.8.1, with signal RMSE and ICC recomputed at each level to assess reproducibility and noise resilience.

## Result

### 3.1 Ablation Results

Architectural ablation was performed from the default PACE configuration at SNR 25 dB across five different seeds. The summarized ablation heatmap (Figure 3) showed that none of the alternative configurations provided a consistent improvement over the default PACE architecture.

Adding $b_0$ to the spatial pathway raised reconstruction RMSE across all IVIM parameters, particularly for $f_{int}$ (0.132 to 0.153) and $D_{par}$ (0.206 * $10^{-3}$ to 0.236 * $10^{-3}$ mm$^2$/s). Alternatively, adding neighboring IVIM signals as spatial input also increased RMSE in four out of five IVIM parameters, with $f_{int}$ (0.132 to 0.141) and $D_{par}$ (0.206 * $10^{-3}$ to 0.218 * $10^{-3}$ mm$^2$/s) as the most affected parameters. Replacing T1 and FLAIR inputs with neighboring IVIM signals while keeping the attention mechanism intact increased parameter RMSE on four of five IVIM parameters: $D_{par}$ increased to 0.226 * $10^{-3}$ mm$^2$/s, $D_{mv}$ to 0.0704, $f_{int}$ to 0.147, and $f_{mv}$ 0.0242.

Replacing cross-attention with concatenation also similarly decreased performance, with $f_{int}$ parameter RMSE increased 9.6% from 0.132 to 0.145. Similarly, closing the selective gating increased $D_{par}$ RMSE ($0.206 * 10^{-3}$ to $0.227*10^{-3}$ mm$^2$/s) and $f_{int}$ RMSE (0.132 to 0.147) by approximately 9.7% and 11.4% respectively.

Switching off $\mathcal{L}_{CE}$ did not produce any noticeable RMSE changes with all the parameter RMSEs remaining stable, while increasing the neighborhood size from 3 x 3 to 5 x 5 increased RMSE on all five parameters (Figure 3). Overall, $D_{par}$ and $f_{int}$ were consistently the most sensitive parameters with respect to changes in spatial architecture.

Lesion CNR RMSE result was mixed: most ablations worsened $D_{par}$ and $D_{mv}$ 's CNR while slightly improved $f_{int}$, with mixed effects on $D_{int}$ and $f_{mv}$ (Figure S2).

**3.2 Anatomy-Weighted Spatial Prior**

Introducing spatial prior from 3x3 neighborhood anatomical patch to PACE universally reduced the parameter RMSE across all 5 IVIM metrics (Figure 4, top row). The largest RMSE reductions were observed for $D_{par}$ ($0.206 * 10^{-3}$ to $0.166 * 10^{-3}$ mm$^2$/s), fint (0.132 to 0.121), and $f_{mv}$ (0.0238 to 0.0202), with smaller improvements for $D_{int}$ ($0.702 * 10^{-3}$ to $0.694 * 10^{-3}$ mm$^2$/s) and $D_{mv}$ (0.0680 to 0.0636).

By contrast, spatial prior's impact on lesion CNR RMSE was parameter dependent. CNR error decreased noticeably for $D_{par}$ (0.229 to 0.0988) and $D_{int}$ (0.0957 to 0.0758) and slightly increased for $D_{mv}$ (0.112 to 0.120) and $f_{int}$ (0.0715 to 0.0963). The corresponding WMH distribution between 3x3 prediction and ground truth (Figure 4, bottom row) showed a similar parameter-dependent shift.

Increasing the anatomy-weighted neighborhood from 3x3 to 5x5 produced mixed effects. Parameter RMSE decreased further for $D_{par}$, $D_{mv}$ and $f_{mv}$ but increased for $D_{int}$ and $f_{int}$ despite the extra information. This trade-off could be seen in the pronounced shift in WMH distribution for $f_{int}$ for 5x5 neighborhoods (Figure 4, bottom row).

**3.3 Model Comparison**

**3.3.1 Synthetic Comparison**

Four neural net methods (PACE, PACE with spatial prior, DNN, IVIM-3C), and three conventional methods LSQ, NNLS, and MAP were evaluated across SNRs 20 to 35 dB using the comparison metrics discussed in section 2.11 (Figure 5).

Averaged across all noise levels, PACE with spatial prior achieved the lowest error for four of five parameters: $D_{par}$ ($0.175* 10^{-3}$ mm$^2$/s), $D_{int}$ ($0.669 * 10^{-3}$ mm$^2$/s), $f_{int}$ (0.126), and $f_{mv}$ (0.0201), with default PACE as a close second for the same parameters. The performance gap over the best non-PACE method was most pronounced for $D_{par}$ (compared to MAP's $0.222 * 10^{-3}$ mm$^2$/s) and $f_{int}$ (compared to DNN's 0.188). IVIM-3C achieved the lowest $D_{mv}$ RMSE at $56.1* 10^{-3}$ mm$^2$/s, with PACE with spatial prior ranking second (60.9 * $10^{-3}$ $mm^2/s$). Performance on the signal degenerated subset was broadly similar to the full test set, with the PACE retaining the lowest RMSE for four out five parameters except for $D_{mv}$ (Table S1).

Lesion CNR result showed a mixed pattern with no clear winners. MAP attained the lowest CNR error for $D_{par}$ (0.121), NNLS for $D_{int}$ (0.0242), LSQ for $D_{mv}$ and $f_{mv}$ (0.104 and 0.0746), and default PACE for $f_{int}$ (0.136). Despite ranking first for $f_{int}$, PACE placed fourth and fifth for the remaining parameters. Adding spatial priors to PACE improved lesion CNR for $D_{par}$, $D_{int}$, and $D_{mv}$ but reduced fidelity for $f_{int}$ and $f_{mv}$, while shifting the distributions of $D_{int}$ and $f_{int}$ closer to the ground truth profiles (Figure 6).

**3.3.2 In-vivo comparison**

Due to the absence of ground truth, in-vivo performance across candidate models was evaluated using signal RMSE against the measured signal (Figure 7). Overall, the neural network methods outperformed the conventional LSQ, NNLS, and MAP across all SNRs tested, with the largest difference observed at lower SNR levels. PACE with spatial prior achieved the lowest signal RMSE averaged across all tissue types and noise levels, with signal RMSE at 0.0639 for 20 dB and 0.0529 for 35 dB. The default PACE and DNN closely trailed closely with similar performances, followed by IVIM-3C (Figure 7). The advantage of spatial information was most noticeable at lower SNRs but diminished as SNR increased. Reproducibility across five training seeds was evaluated for the PACE variants and DNN configurations. All three methods have good-to-excellent reproducibility, with ICC above 0.8 for all IVIM parameters (Figure S3). DNN were the most consistent except for $f_{int}$ (0.841 against PACE's 0.969), while the PACE models showed more variability with across different seed initializations.

**3.3.3 Hybrid brain**

Figure 8 shows the predicted parameter maps against the hybrid brain ground truth at 25 dB, with their respective performance reported in Table S2. PACE with spatial prior achieved the lowest RMSE for four out of five IVIM parameters: $D_{par}$ (0.155 x $10^{-3}$ $mm^2/s$), $D_{int}$ (0.716 x 10-3 mm^2/s), $f_{int}$ (0.111), and $f_{mv}$ (0.0235). The largest relative improvements over DNN were $D_{par}$ (0.155 * $10^{-3}$ $mm^2/s$ vs 0.237 * $10^{-3}$ $mm^2/s$) and $f_{int}$ (0.111 vs 0.207). IVIM-3C achieved the lowest $D_{mv}$ RMSE (58.0 x $10^{-3}$ $mm^2/s$) and the lowest overall signal RMSE (0.0231), despite having higher RMSE than PACE with prior on the remaining four parameters.

## Discussion

In this work, we introduced PACE, a multimodal physics-informed neural network designed to leverage spatial context for IVIM fitting. Furthermore, we incorporated a spatial prior to test whether anatomy-weighted signal could effectively resolve IVIM degeneracy. Overall, the integration of spatial inputs and priors significantly improved IVIM parameter reconstruction accuracy, with moderate CNR trade-offs.

### 4.1 IVIM degeneracy and parameter identifiability for the intermediate component

The intermediate component is particularly vulnerable to parameter degeneracy. At higher $b$ regimes ($b$>700 s/mm$^2$), the parenchymal component dominates, anchoring $D_{par}$ estimation. At the lower $b$ regimes (b<100 s/mm$^2$), data were densely sampled with higher SNRs, allowing $D_{mv}$ and $f_{mv}$ to be effectively constrained. In the intermediate regimes, all three exponential contributions coexist simultaneously without asymptotic limits, leaving intermediate parameters underdetermined. Thus, different combinations of $D_{int}$ and $f_{int}$ could satisfy the remaining signal equally, leading to degeneracy behavior under sparse sampling conditions. PACE variants showed lower $D_{int}$ and $f_{int}$ RMSE than the non-PACE models in our synthetic evaluation (Figure 5), suggesting that spatial information provided additional benefits for resolving the less identifiable intermediate components.

### 4.2 Architectural ablation and spatial context filtering

The ablation experiments demonstrated that PACE's improvements benefited from selective integration of spatial information rather than simply adding neighboring signal information to the model (Figure 3). First, the type of spatial input matters, so extra information did not necessarily improve IVIM fitting. Adding the $b_0$ or full neighborhood signal (IVIM_nb) to the existing spatial inputs introduced contrast artifacts and decreased fitting performance. Replacing T1w and FLAIR with IVIM_nb also degraded performance, suggesting that anatomical T1w and FLAIR provide complementary spatial context beyond neighboring IVIM signals alone.

Second, spatial data filtering matters. Removing the attention weighting or the spatial gate mechanism both reduced fitting performance. Cross attention controls which spatial features contribute to the spatial context, while the parameter gating mechanism regulates how strongly the spatial context influenced the IVIM pathway. Equation (4) explicitly enforced these constraints, allowing the network to draw on spatial context only if doing so lowers the IVIM loss. Enabling $\mathcal{L}_{CE}$ did not significantly impact the fitting (Figure 3) but improved CNR agreement for $D_{int}$, $D_{mv}$ and $f_{int}$ (Figure S2) and was therefore retained.

The largest ablation effects were observed for $D_{par}$ and $f_{int}$, the two parameters with the greatest separations between tissue class in the ground truth distributions (Figure 3). This suggests that spatial anatomy is most informative when anatomical tissue contrast corresponds to the differences in the underlying IVIM parameters. Conversely, parameters with limited tissue-specific contrast may receive less benefit. The 3x3 neighborhood therefore appeared to have sufficient anatomical context to associate the center IVIM signal with local tissue profiles. Expanding the window likely introduced less relevant anatomical contrast and did not improve IVIM fitting performance (Figure 3).

Across the architectural ablations, lesion CNR RMSE showed a consistent trade-off between $D_{par}$ /$D_{mv}$ and $f_{int}$ contrast accuracy, hinting that the current spatial-context architecture may not simultaneously preserve diffusivity and fractional contrast (Figure S2).

### 4.3 The Effect of Anatomy-Weighted Spatial Prior

Building on Kaandorp and Zijlstra's spatial approaches[33], we extended to incorporate neighboring signals using anatomy-weighted information. In self-supervised setting, direct incorporation of noisy neighboring voxel signals added little benefits, as observed in Kaandorp[33] as well as in our ablation study. Our anatomy-weighted spatial prior addressed this limitation by first weighting neighborhood signals according to anatomical similarity before incorporating them into the loss.

By scaling signal contribution based on T1w and FLAIR distance, the effective target becomes an anatomy-weighted average favoring voxels most similar to the center (Section 2.6). This anatomy-conditioned $\mathcal{L}_{spatial}$ thus provides an additional constraint without directly prescribing an IVIM parameter ground truth or tissue labels.

The PACE model with 3x3 spatial prior reduced RMSE across all five IVIM parameters relative the default configuration (Figure 4), with the overall performance gain established for both the synthetic and hybrid brain test sets (Figure 5 and Table S1). However, the 5x5 variant's mixed performance implies that larger neighborhoods may incorporate less relevant local information.

The gain in parameter accuracy came with a lesion contrast trade-off. The spatial prior improved CNR agreement for the diffusivities but moderately worsened $f_{int}$ and $f_{mv}$, suggesting that local signal regularization may also introduce spatial bias (Figure 6).

### 4.4 Hybrid Brain and In-vivo Performance

In the hybrid brain at 25 dB, synthetically trained PACE variants outperformed baseline models on realistic brain anatomy, despite being trained on anatomical patches from different subjects (Section 3.4). Interestingly, despite IVIM-3C having the lowest brain-averaged signal RMSE, PACE with spatial prior produced the lowest RMSE for four of five IVIM parameters (Table S2). This demonstrates that superior signal reconstruction does not always guarantee better IVIM parameter accuracy.

The in-vivo results thus require conservative interpretation, since the signal RMSE is the surrogate benchmark in the absence of ground truth. PACE with spatial prior maintained the lowest signal RMSE across all tested noise levels (section 3.3.2), with its advantage most prominent at lower SNRs (Figure 7). While this hints that spatial information may provide more stability for signal reconstruction, whether this trend extends to in-vivo parameter RMSE remains uncertain.

ICC reproducibility remained good to excellent for all PACE variants, although generally lower than DNN, suggesting that spatial information may introduce greater model variability.

### 4.5 Limitations

Our method comes with several limitations. Incorporating spatial contexts increased computational overhead and depended on accurate anatomical segmentation and registration. Tissue label accuracy also contributed to training through classification loss, although it had little effect on performance. Performance may further depend on the resolution of the input modalities, while the impact of partial volume effects was not accounted for. While PACE variants uniformly improved parameter RMSE, they showed mixed results for lesion contrast fidelity, suggesting that the current spatial-context architecture could not uniformly preserve diffusivity and fractional contrasts. Thus, further investigation is needed to determine the tradeoffs between incorporating spatial information and tissue contrast. Additionally, interactions between anatomy-weighted prior and other PACE ablations were not assessed. Furthermore, training and evaluation were performed on a single cohort using one acquisition protocol. The benefit of spatial information depends on the relationship between anatomical contrast and IVIM parameter distributions, which may vary across different protocols and disease cohorts. Finally, the synthetic parameter ranges were derived from LSQ estimates and may thus inherit their bias. The relatively sparse sampling in the intermediate b-value regimes also limited estimation accuracy of the intermediate diffusion parameters.

## CONCLUSION

In this work, we proposed the **P**hysics-**A**nchored **C**ross-attention **E**mbedding (PACE) network, an advanced multimodal physics-informed network for three-component IVIM fitting. By leveraging distinct T1w and FLAIR signatures from different tissue types, PACE successfully improved reconstruction accuracy compared to DNNs and conventional models. The anatomy-weighted spatial prior provides a novel approach to reduce signal degeneracies by employing local tissue information, enhancing parameter reconstruction across all five parameters at the moderate cost of perfusion and interstitial lesion contrast. These findings demonstrate a promising proof of concept for utilizing anatomically guided, physics-informed approaches in future IVIM research.

## ACKNOWLEDGEMENTS:

This study was supported by NIH’s grant R01AG077156.

## DATA AVAILABILITY STATEMENT

The synthetic data and fitting approaches used in this paper is available at https://github.com/gavtoski/PACE_IVIM

**REFERENCE**

1. Le Bihan D. What can we see with IVIM MRI? Neuroimage. 2019;187:56–67. doi: 10.1016/j.neuroimage.2017.12.062.
2. Gurney-Champion OJ, Klaassen R, Froeling M, Barbieri S, Stoker J, Engelbrecht MRW, Wilmink JW, Besselink MG, Bel A, van Laarhoven HWM, Nederveen AJ. Comparison of six fit algorithms for the intra-voxel incoherent motion model of diffusion-weighted magnetic resonance imaging data of pancreatic cancer patients. PLoS One. 2018;13(4):e0194590. doi: 10.1371/journal.pone.0194590.
3. Iima M. Perfusion-driven Intravoxel Incoherent Motion (IVIM) MRI in Oncology: Applications, Challenges, and Future Trends. Magn Reson Med Sci. 2021;20(2):125–38. doi: 10.2463/mrms.rev.2019-0124.
4. Kaandorp MPT, Barbieri S, Klaassen R, van Laarhoven HWM, Crezee H, While PT, Nederveen AJ, Gurney-Champion OJ. Improved unsupervised physics-informed deep learning for intravoxel incoherent motion modeling and evaluation in pancreatic cancer patients. Magnetic Resonance in Med. 2021;86(4):2250–65. doi: 10.1002/mrm.28852.
5. Le Bihan D, Ichikawa S, Motosugi U. Diffusion and Intravoxel Incoherent Motion MR Imaging–based Virtual Elastography: A Hypothesis-generating Study in the Liver. Radiology. 2017;285(2):609–19. doi: 10.1148/radiol.2017170025.
6. Li YT, Cercueil J-P, Yuan J, Chen W, Loffroy R, Wáng YXJ. Liver intravoxel incoherent motion (IVIM) magnetic resonance imaging: a comprehensive review of published data on normal values and applications for fibrosis and tumor evaluation. Quant Imaging Med Surg. 2017;7(1):59–78. doi: 10.21037/qims.2017.02.03.
7. Podgórska J, Pasicz K, Zagórowicz E, Mróz A, Gołębiewski B, Kuś P, Jasieniak J, Skrzyński W, Wieszczy P, Kukołowicz P, Cieszanowski A. Intravoxel incoherent motion MRI in evaluating inflammatory activity in ulcerative colitis: a pilot study. Acta Radiol. 2021;62(4):439–46. doi: 10.1177/0284185120931689.
8. Cho GY, Moy L, Kim SG, Baete SH, Moccaldi M, Babb JS, Sodickson DK, Sigmund EE. Evaluation of breast cancer using intravoxel incoherent motion (IVIM) histogram analysis: comparison with malignant status, histological subtype, and molecular prognostic factors. Eur Radiol. 2016;26(8):2547–58. doi: 10.1007/s00330-015-4087-3.
9. Alomair OI, Alghamdi SA, Abujamea AH, Alfifi AY, Alashban YI, Kurniawan ND. Investigating the Role of Intravoxel Incoherent Motion Diffusion-Weighted Imaging in Evaluating Multiple Sclerosis Lesions. Diagnostics. 2025;15(10):1260. doi: 10.3390/diagnostics15101260.

10. Paschoal AM, Leoni RF, dos Santos AC, Paiva FF. Intravoxel incoherent motion MRI in neurological and cerebrovascular diseases. NeuroImage : Clinical. 2018;20:705–14. doi: 10.1016/j.nicl.2018.08.030.
11. Suo S, Cao M, Zhu W, Li L, Li J, Shen F, Zu J, Zhou Z, Zhuang Z, Qu J, Chen Z, Xu J. Stroke assessment with intravoxel incoherent motion diffusion-weighted MRI. NMR Biomed. 2016;29(3):320–8. doi: 10.1002/nbm.3467.
12. Thiel MM, Freeze WM, Jong JJ, Ramakers IHGB, Backes WH, Jansen JFA. Interstitial fluid as a proxy of glymphatic dysfunction in patients with cognitive impairment: The necessity of three-directional intravoxel incoherent motion. Alzheimer's & Dementia. 2021;17(S4):e052105. doi: 10.1002/alz.052105.
13. Gomolka RS, Hablitz LM, Mestre H, Giannetto M, Du T, Hauglund NL, Xie L, Peng W, Martinez PM, Nedergaard M, Mori Y. Loss of aquaporin-4 results in glymphatic system dysfunction via brain-wide interstitial fluid stagnation. Elife. 2023;12:e82232. doi: 10.7554/eLife.82232.
14. Kerkhofs D, Wong SM, Zhang E, Staals J, Jansen JFA, van Oostenbrugge RJ, Backes WH. Baseline Blood-Brain Barrier Leakage and Longitudinal Microstructural Tissue Damage in the Periphery of White Matter Hyperintensities. Neurology. 2021;96(17):e2192–e200. doi: 10.1212/WNL.0000000000011783.
15. van der Thiel MM, Freeze WM, Verheggen ICM, Wong SM, de Jong JJA, Postma AA, Hoff EI, Gronenschild EHBM, Verhey FR, Jacobs HIL, Ramakers IHGB, Backes WH, Jansen JFA. Associations of increased interstitial fluid with vascular and neurodegenerative abnormalities in a memory clinic sample. Neurobiol Aging. 2021;106:257–67. doi: 10.1016/j.neurobiolaging.2021.06.017.
16. Wong SM, Backes WH, Drenthen GS, Zhang CE, Voorter PHM, Staals J, Oostenbrugge RJv, Jansen JFA. Spectral Diffusion Analysis of Intravoxel Incoherent Motion MRI in Cerebral Small Vessel Disease. Journal of Magnetic Resonance Imaging. 2020/04/01;51(4). doi: 10.1002/jmri.26920.
17. Wong SM, Zhang CE, van Bussel FCG, Staals J, Jeukens CRLPN, Hofman PAM, van Oostenbrugge RJ, Backes WH, Jansen JFA. Simultaneous investigation of microvasculature and parenchyma in cerebral small vessel disease using intravoxel incoherent motion imaging. NeuroImage : Clinical. 2017;14:216–21. doi: 10.1016/j.nicl.2017.01.017.
18. Voorter PHM, Backes WH, Gurney-Champion OJ, Wong S-M, Staals J, van Oostenbrugge RJ, van der Thiel MM, Jansen JFA, Drenthen GS. Improving microstructural integrity, interstitial fluid, and blood microcirculation images from multi-b-value diffusion MRI using physics-informed neural networks in cerebrovascular disease. Magnetic Resonance in Med. 2023;90(4):1657–71. doi: 10.1002/mrm.29753.
19. van Dinther M, Voorter PHM, Zhang E, van Kuijk SMJ, Jansen JFA, van Oostenbrugge RJ, Backes WH, Staals J. The neurovascular unit and its correlation with cognitive performance in patients with cerebral small vessel disease: a canonical correlation analysis approach. GeroScience. 2024;46(5):5061–73. doi: 10.1007/s11357-024-01235-8.
20. Scalco E, Rizzo G, Mastropietro A. The quantification of IntraVoxel incoherent motion – MRI maps cannot preserve texture information: An evaluation based on simulated and in-vivo images. Computers in Biology and Medicine. 2023;154:106495. doi: 10.1016/j.compbiomed.2022.106495.

21. Kaandorp MPT, Zijlstra F, Federau C, While PT. Deep learning intravoxel incoherent motion modeling: Exploring the impact of training features and learning strategies. Magnetic Resonance in Med. 2023;90(1):312–28. doi: 10.1002/mrm.29628.
22. Koopman T, Martens R, Gurney-Champion OJ, Yaqub M, Lavini C, de Graaf P, Castelijns J, Boellaard R, Marcus JT. Repeatability of IVIM biomarkers from diffusion-weighted MRI in head and neck: Bayesian probability versus neural network. Magnetic Resonance in Med. 2021;85(6):3394–402. doi: 10.1002/mrm.28671.
23. Barbieri S, Gurney-Champion OJ, Klaassen R, Thoeny HC. Deep learning how to fit an intravoxel incoherent motion model to diffusion-weighted MRI. Magnetic Resonance in Med. 2020;83(1):312–21. doi: 10.1002/mrm.27910.
24. Troelstra MA, Van Dijk A-M, Witjes JJ, Mak AL, Zwirs D, Runge JH, Verheij J, Beuers UH, Nieuwdorp M, Holleboom AG, Nederveen AJ, Gurney-Champion OJ. Self-supervised neural network improves tri-exponential intravoxel incoherent motion model fitting compared to least-squares fitting in non-alcoholic fatty liver disease. Frontiers in Physiology. 2022;13. doi: 10.3389/fphys.2022.942495.
25. Faiyaz A, Doyley M, Schifitto G, Zhong J, Uddin MN. Single-shell NODDI using dictionary-learner-estimated isotropic volume fraction. NMR Biomed. 2022;35(2):e4628. Epub 20211012. doi: 10.1002/nbm.4628. PubMed PMID: 34642974.
26. Faiyaz A, Doyley MM, Schifitto G, Uddin MN. Artificial intelligence for diffusion MRI-based tissue microstructure estimation in the human brain: an overview. Front Neurol. 2023;Volume 14 - 2023. doi: 10.3389/fneur.2023.1168833.
27. Qiu X, Uddin MN, Wang L, Tyrell A, Faiyaz A, Hoang N, Zhuang Y, Tivarus ME, Zhong J, Weber MT, Schifitto G. A multimodal imaging–based integrative framework for HIV-associated cognitive impairment and treatment response. Frontiers in Neuroscience. 2026;Volume 20 - 2026. doi: 10.3389/fnins.2026.1775047.
28. Manjón JV, Coupé P. Frontiers | volBrain: An Online MRI Brain Volumetry System. Frontiers in Neuroinformatics. 2016/07/27;10. doi: 10.3389/fninf.2016.00030.
29. Kamraoui RA, Ta V-T, Tourdias T, Mansencal B, Manjon JV, Coupé P. DeepLesionBrain: Towards a broader deep-learning generalization for multiple sclerosis lesion segmentation. Medical Image Analysis. 2022;76:102312. doi: 10.1016/j.media.2021.102312.
30. Vaswani A, Shazeer N, Parmar N, Uszkoreit J, Jones L, Gomez AN, Kaiser Łu, Polosukhin I, editors. Attention is All you Need2017 2017: Curran Associates, Inc.
31. Rodriguez-Molares A, Rindal OMH, D'Hooge J, Masoy S-E, Austeng A, Lediju Bell MA, Torp H. The Generalized Contrast-to-Noise Ratio: A Formal Definition for Lesion Detectability. IEEE Transactions on Ultrasonics, Ferroelectrics, and Frequency Control. 2020;67(4):745–59. doi: 10.1109/tuffc.2019.2956855.
32. Koo TK, Li MY. A Guideline of Selecting and Reporting Intraclass Correlation Coefficients for Reliability Research. Journal of Chiropractic Medicine. 2016;15(2):155–63. doi: 10.1016/j.jcm.2016.02.012.
33. Kaandorp MPT, Zijlstra F, Karimi D, Gholipour A, While PT. Incorporating spatial information in deep learning parameter estimation with application to the intravoxel incoherent motion model in diffusion-weighted MRI. Medical Image Analysis. 2025;101:103414. doi: 10.1016/j.media.2024.103414.

**LIST OF TABLES**

| Class | $D_{par}$ ($10^{-3}$ mm²/s) | $D_{int}$ ($10^{-3}$ mm²/s) | $D_{mv}$ ($10^{-3}$ mm²/s) | $f_{int}$ | $f_{mv}$ |
|---|---|---|---|---|---|
| GM | 0.65 ± 0.18 | 2.71 ± 0.71 | 96.3 ± 57.8 | 0.28 ± 0.12 | 0.038 ± 0.041 |
| NAWM | 0.59 ± 0.17 | 2.35 ± 0.61 | 91.0 ± 51.9 | 0.25 ± 0.11 | 0.021 ± 0.020 |
| WMH | 0.82 ± 0.23 | 2.49 ± 0.67 | 78.1 ± 43.0 | 0.36 ± 0.15 | 0.016 ± 0.015 |

**Table 1:** Tissue-specific distributions used for synthetic data generation, as taken from the LSQ fitting of 52 reserved subjects. These results established the tissue-class specific parameter ranges from which synthetic parameters were sampled using log normal distribution to create the synthetic ground truth for our data set. Note that $D_{par}$ and $f_{int}$ have the most distinct distribution across the three different tissue types. Twenty percent of the patches were planted as cross-tissue degenerate pairs whose IVIM attenuation signals differ by less than 0.2% in root mean square (RMS) across b-values.

| Settings | PACE variants | DNN | IVIM3C |
| --- | --- | --- | --- |
| Inputs | IVIM signal<br>T1w, FLAIR neighborhood patches | IVIM signal | IVIM signal |
| Losses | $\mathcal{L}_{IVIM} + \alpha_{spatial}\mathcal{L}_{spatial} + \alpha_{CE}\mathcal{L}_{CE},$ | $\mathcal{L}_{IVIM}$ | $\mathcal{L}_{IVIM}$ |
| Bounds, diffusivities ($10^{-3}$ mm²/s) | $D_{par}$: [0.1, 1.5], $D_{int}$ [1.5, 4.0], $D_{mv}$ [4, 200] | same as PACE | same as PACE |
| Bounds, IVIM fractions | $f_{int}$ [0, 0.4], $f_{mv}$ [0, 0.2] | same as PACE | same as PACE |
| Optimizer | AdamW, weight decay $1 \times 10^{-5}$ | same as PACE | Adam |
| Learning rate | $2.5 \times 10^{-3}$ spatial encoders, attention, decoders; $5 \times 10^{-4}$ IVIM encoder, gates; $1 \times 10^{-4}$ physics heads | same as PACE | $3 \times 10^{-5}$ |
| Learning schedule | Warm up to epoch 5, constant to 20, cosine decay | same as PACE | Reduce on plateau |

**Table 2**: Hyper-parameter settings for training neural networks. PACE variants and DNN are the different ablations of the same network, sharing most of the hyper-parameter settings except for the spatial-related modules. IVIM-3C were kept to the published default setting, with the bounds modified to match PACE and DNN. *Abbreviation: PACE - **P**hysics-**A**nchored **C**ross-Attention **E**mbedding, DNN – **D**eep **N**eural **N**et, IVIM3C- Published IVIM deep net by Voorter et al.*[18].

**LIST OF FIGURES**

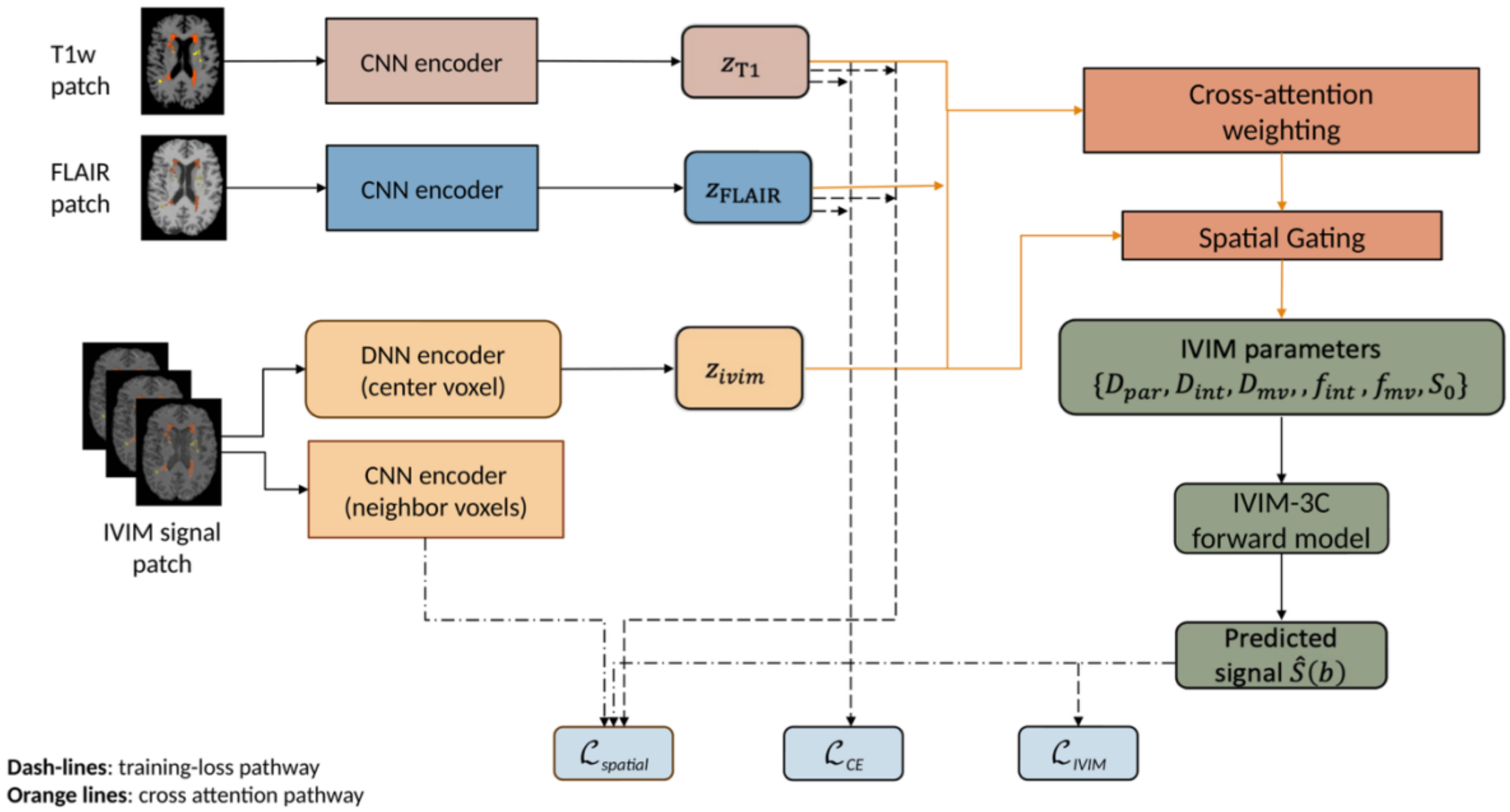

***Figure 1.*** *Schematic of PACE (**P**hysics-**A**nchored **C**ross-Attention **E**mbedding) network. T1w and FLAIR patches are encoded separately and fused with the signal IVIM embedding of the center voxel through cross-attention mechanism. Spatial gating filters outputs from the cross-attention block for helpful spatial contribution before passing IVIM parameters to the forward model for signal reconstruction. For the default PACE with prior, training uses IVIM signal reconstruction loss* $\mathcal{L}_{IVIM}$ *and two auxiliary losses* $\mathcal{L}_{CE}$ *for tissue classification and* $\mathcal{L}_{spatial}$ *anatomy-weighted signal consistency. The training loss pathway are denoted as dashed lines, with the mathematical formulations detailed in section 2.6 and 2.7.*

$$d_{ij}^2 = \left(T1_i - T1_j\right)^2 + \left(\mathrm{FLAIR}_i - \mathrm{FLAIR}_j\right)^2$$

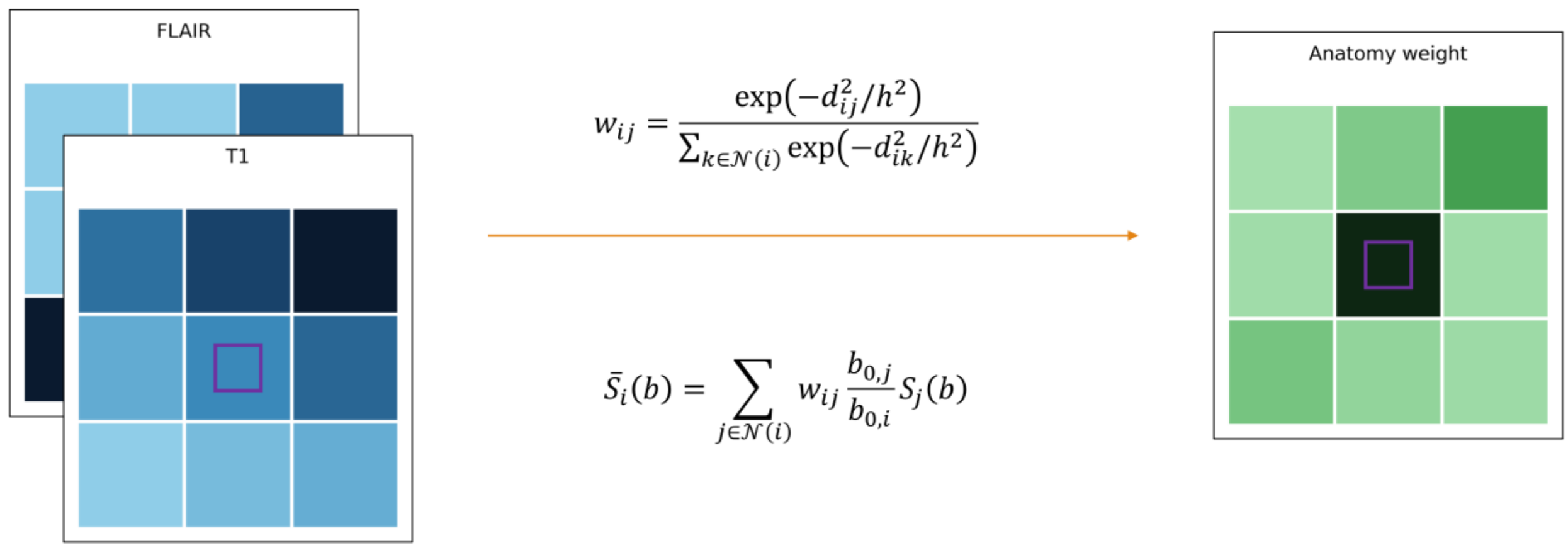

$$w_{ij} = \frac{\exp\left(-d_{ij}^2/h^2\right)}{\sum_{k\in\mathcal{N}(i)} \exp\left(-d_{ik}^2/h^2\right)}$$

$$\bar{S}_i(b) = \sum_{j\in\mathcal{N}(i)} w_{ij} \frac{b_{0,j}}{b_{0,i}} S_j(b)$$

***Figure 2:*** *Anatomy-weighted spatial prior computation. The top equations explained how anatomical similarity distance* $d_{ij}$ *between the target voxel (marked by the purple square) are computed from the T1w and FLAIR intensities. The middle equation showed how neighbor anatomical similarity is scaled and normalized into neighborhood weight* $w_{ij}$*, giving greater leverage to neighbor voxels with similar tissue properties. The last equation showed how the effective patch signal* $\tilde{S}(b)$ *is computed using the individual weight* $w_{ij}$*, the normalized neighborhood IVIM signal* $S_j(b)$*, and neighbor/center amplitude ratio* $\frac{b_{0,j}}{b_{0,i}}$ *.*

Architecture ablation: signal and parameter RMSE

| | Spatial input | Fusion | Gate | CE | Patch | Signal | $D_{par}$ ($10^{-3}$ mm$^2$/s) | $D_{int}$ ($10^{-3}$ mm$^2$/s) | $D_{mv}$ ($10^{-3}$ mm$^2$/s) | $f_{int}$ | $f_{mv}$ |
|---|---|---|---|---|---|---|---|---|---|---|---|
| PACE | T1+FLAIR | attention | yes | yes | 3x3 | 0.0138 | 0.206 | 0.702 | 68.0 | 0.132 | 0.0238 |
| PACE, with b0 | T1+FLAIR+b0 | attention | yes | yes | 3x3 | 0.0140 | 0.236 | 0.716 | 71.1 | 0.153 | 0.0250 |
| PACE-concat | T1+FLAIR | concatenate | yes | yes | 3x3 | 0.0139 | 0.226 | 0.704 | 70.4 | 0.145 | 0.0240 |
| PACE, gates closed | T1+FLAIR | attention | no | yes | 3x3 | 0.0140 | 0.227 | 0.708 | 70.5 | 0.147 | 0.0245 |
| PACE, no CE | T1+FLAIR | attention | yes | no | 3x3 | 0.0137 | 0.200 | 0.707 | 66.9 | 0.131 | 0.0232 |
| PACE, 5x5 patch | T1+FLAIR | attention | yes | yes | 5x5 | 0.0139 | 0.222 | 0.709 | 68.9 | 0.144 | 0.0245 |
| PACE, neighbour input | T1+FLAIR+IVIM_nb | attention | yes | yes | 3x3 | 0.0138 | 0.218 | 0.712 | 68.9 | 0.140 | 0.0237 |
| PACE, no anatomy | IVIM_nb | attention | yes | yes | 3x3 | 0.0137 | 0.226 | 0.703 | 70.4 | 0.147 | 0.0242 |

***Figure 3:*** *Architectural ablation of PACE net at SNR 25 dB, five seedings per arm. The left table summarizes architectural details for ablation model, includes all the studied components: spatial input, fusion strategy, spatial gating selection, tissue classification loss, and patch kernel size. Every ablation differs from the default PACE by one setting. The right table details the RMSE for each of the IVIM parameters, with the PACE default on top as reference. Red-blue shading is relative to default PACE row, red indicating higher errors and blue indicating lower error than the default configuration.*

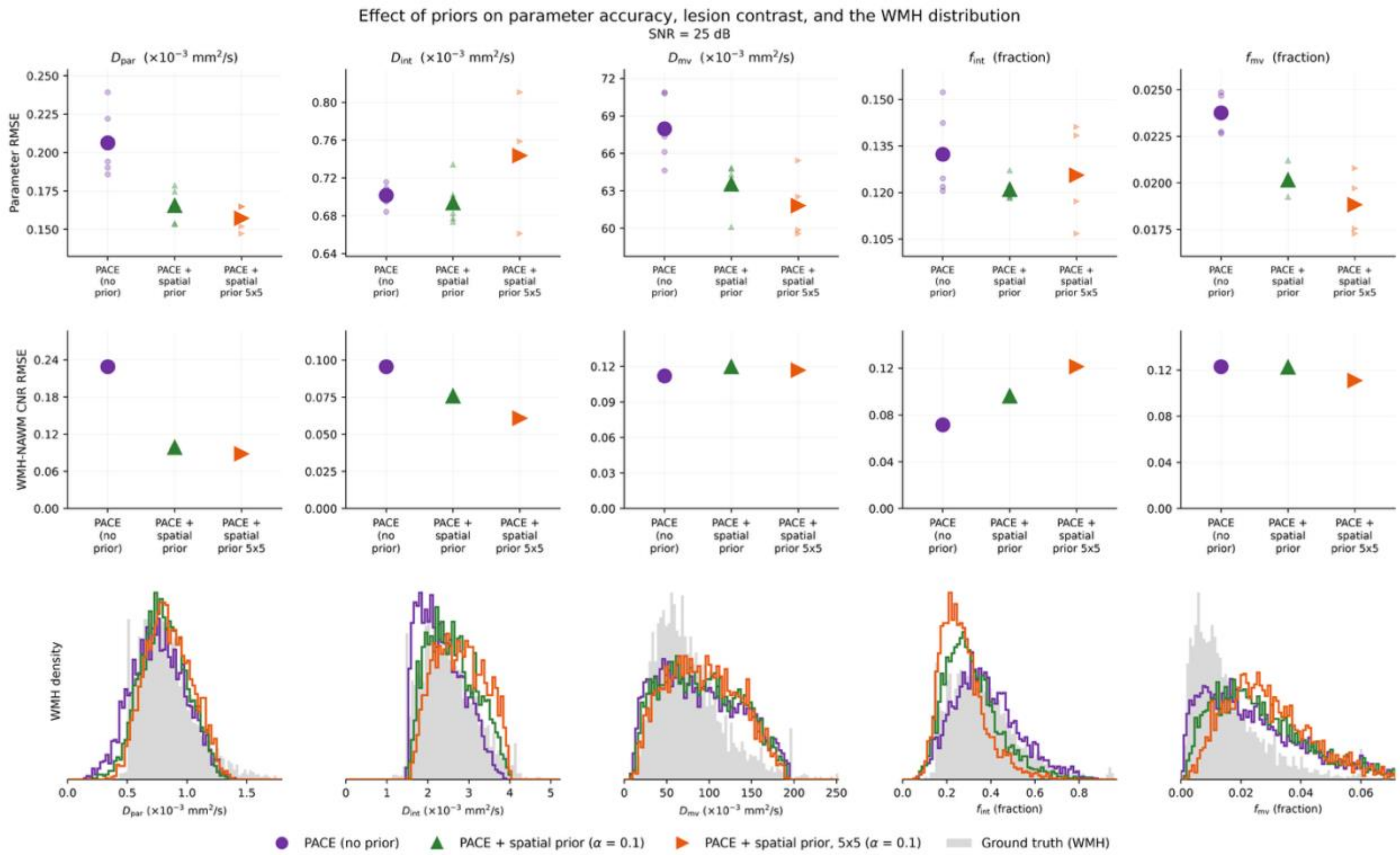

***Figure 4:*** *Effect of spatial prior on IVIM parameter reconstruction accuracy (top), lesion contrast presentation (middle), and WMH distribution (bottom) at SNR 25 dB. Default PACE (purple) is compared with PACE with spatial prior with 3x3 (green) and 5x5 (orange) neighborhoods. Top: The spatial prior 3x3 neighborhood consistently improved accuracy (reducing RMSE) compared to the default PACE, whereas the effect of the 5x5 neighborhood was mixed. Middle: Spatial priors helped with NAWM-WMH CNR agreement, except for* $f_{int}$*, where an opposite trend was observed. Bottom: Predicted WMH distributions between three PACE variants and ground truth, with 3x3 neighborhood spatial prior generally showing the best distribution agreement for the diffusion parameters, while having a negative bias for* $f_{int}$*.*

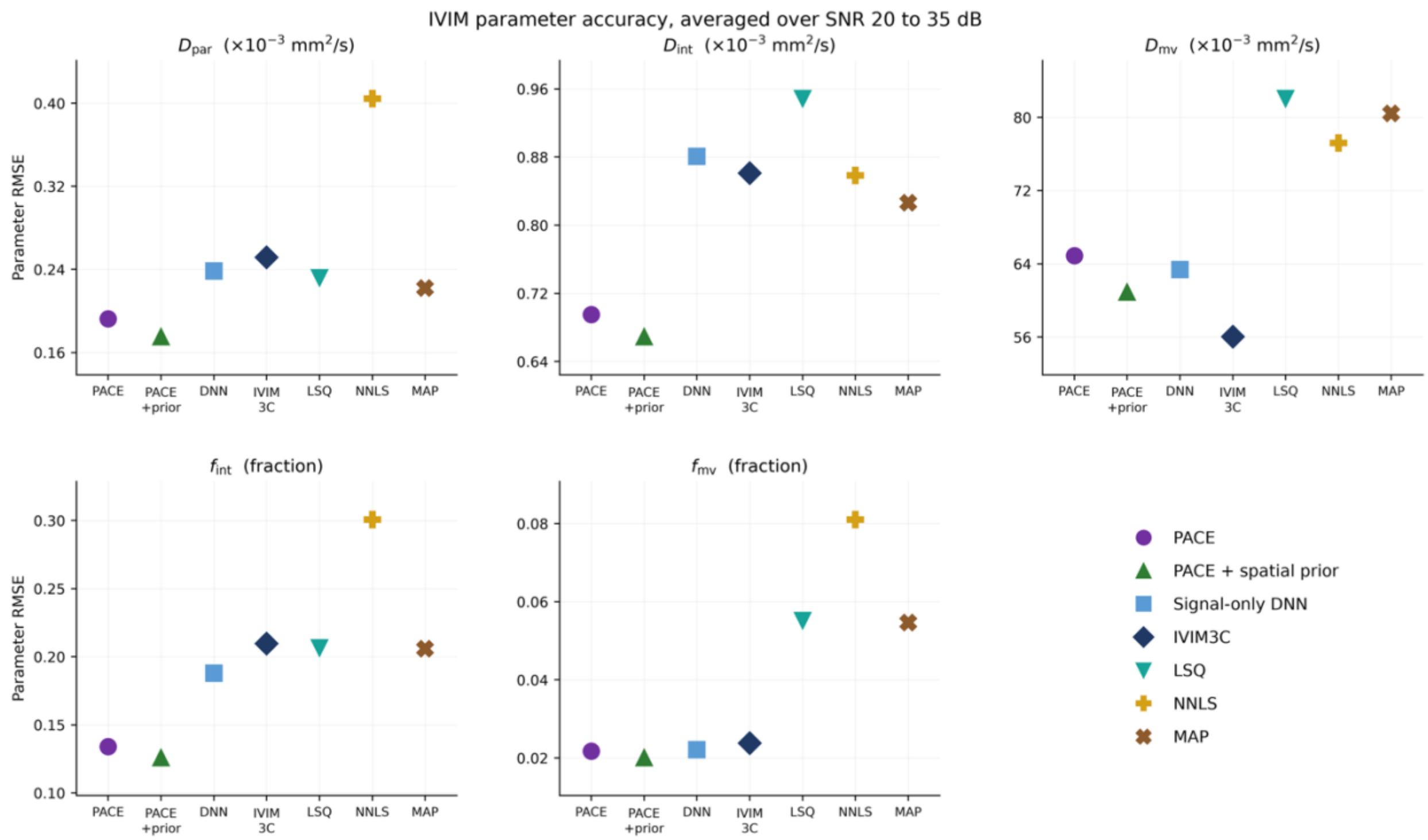

***Figure 5.*** *IVIM parameter reconstruction accuracy across fitting methods, averaged over 20-35 dB. Parameter RMSE is showed for each IVIM parameters, with three diffusivities ($D_{par}$, $D_{int}$, $D_{mv}$) on top with the volume fraction ($f_{int}$, $f_{mv}$) on bottom. The seven candidate methods were color-coded: PACE-purple, PACE with spatial prior-green, DNN-sky blue, LSQ-teal, NNLS-mustard yellow, MAP-brown. Lower RMSE values denotes better parameter reconstruction accuracy. PACE variants have the best and second best performance for $D_{par}$, $D_{int}$, $f_{int}$, and $f_{mv}$ with IVIM-3C claimed best $D_{mv}$ reconstruction. Abbreviation: PACE - **P**hysics-**A**nchored **C**ross-Attention **E**mbedding, DNN – **D**eep **N**eural **N**et, IVIM3C- Published IVIM deep net by Voorter et al.[18], LSQ – **L**east **S**quare, NNLS – **N**on-**N**egative **L**east **S**quare, **MAP** – **M**aximum-**A**-**P**osteriori.*

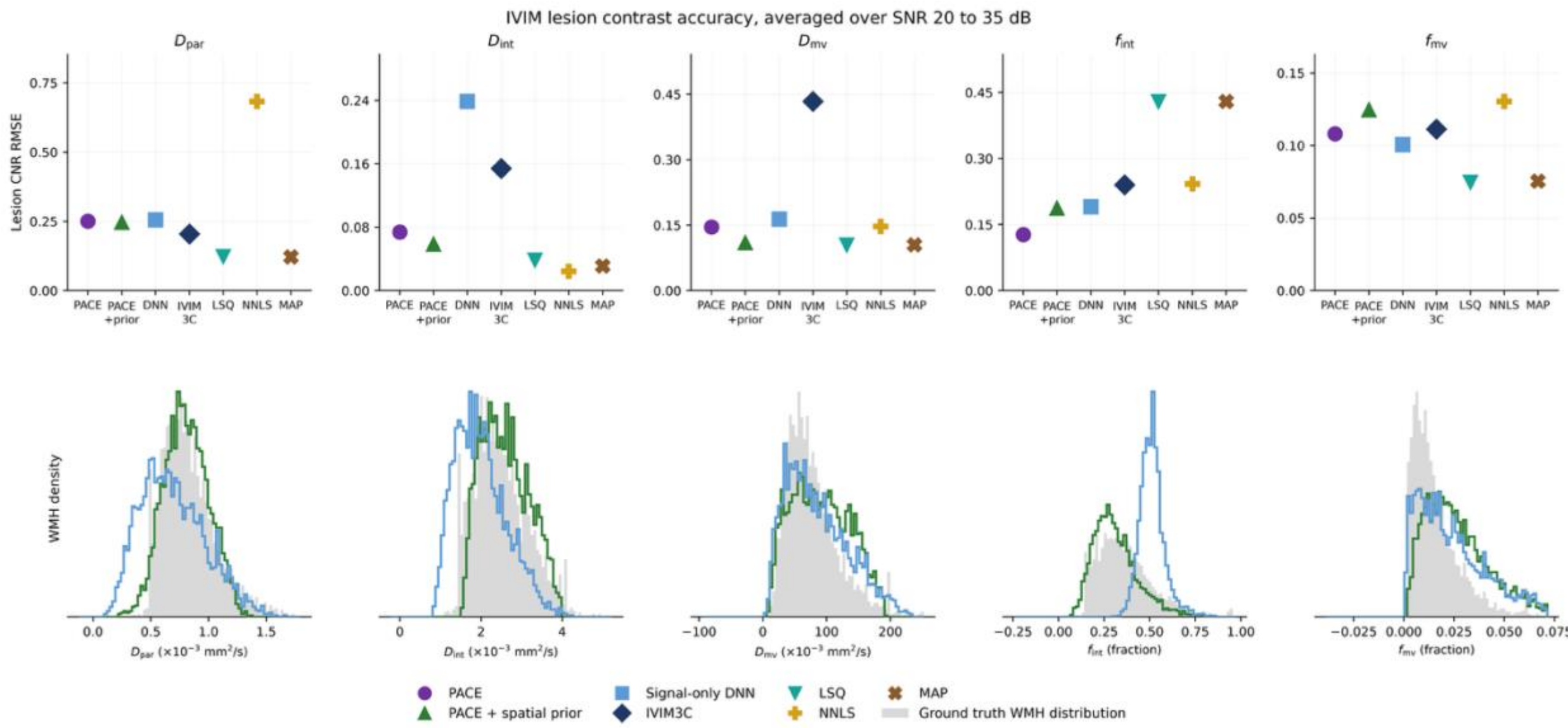

***Figure 6.*** *Lesion contrasts to noise ratio (CNR) performance evaluated across candidate models, averaged over 20-35 dB. Top row: lesion CNR RMSE for the IVIM parameters, with lower values indicating better agreement with the mean ground truth contrast. None of the candidate method was the decisive winner across all parameters. The PACE variants showed more consistent performance across all five IVIM parameters, whereas other methods showed a drop in performance for at least one parameter ($D_{int}$ for DNN, $D_m$v for IVIM3C, $D_{par}$ for LSQ, $f_{int}$ for NNLS and MAP). Adding spatial prior to PACE reduced CNR error for the three diffusivities, while moderately increased errors for $f_{int}$ and $f_{mv}$. Bottom row: Predicted WMH parameter distributions for DNN and PACE variants compared with ground truth. Adding spatial information shifted the distributions of Dint and fint closer to the ground truth profiles.*

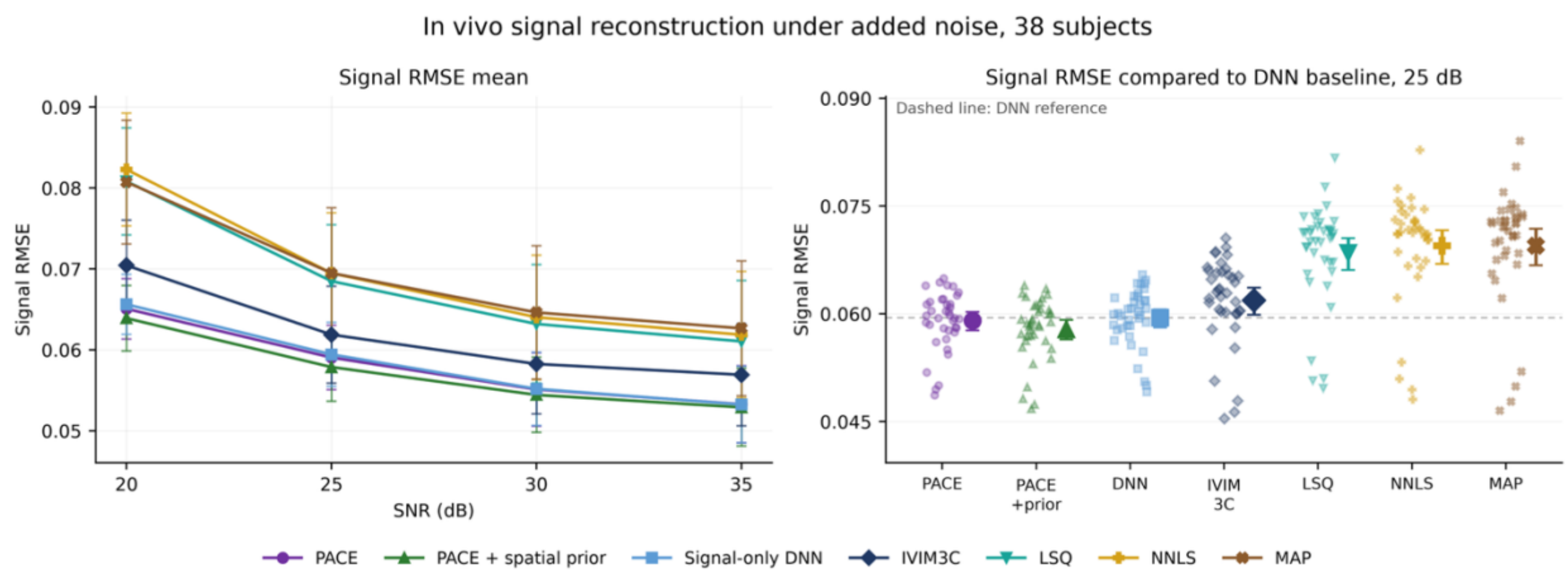

***Figure 7:*** *In-vivo IVIM signal reconstruction performance under added noise across 38 subjects. Left: Mean signal RMSE of seven candidate models (PACE, PACE spatial prior, DNN, IVIM3C, LSQ, NNLS, MAP) across SNRs 20-35 dB. All seven models showed deceasing reconstruction errors as SNR improves as expected. The PACE variants consistently demonstrated the lowest RMSE, with PACE using spatial prior achieving the lowest RMSE across all noise levels. Neural networks methods generally outperformed the conventional fitting methods. Right: Subject level signal RMSE chart at 25 dB, with individual subjects shown as smaller color-coded markers. The horizontal dashed line denotes the mean DNN reference. Abbreviation: PACE -* ***P****hysics-****A****nchored* ***C****ross-Attention* ***E****mbedding, DNN –* ***D****eep* ***N****eural* ***N****et, IVIM3C- Published IVIM deep net by Voorter et al.[18], LSQ –* ***L****east* ***S****quare, NNLS –* ***N****on-Negative* ***L****east* ***S****quare,* ***MAP*** *–* ***M****aximum-****A****-****P****osteriori.*

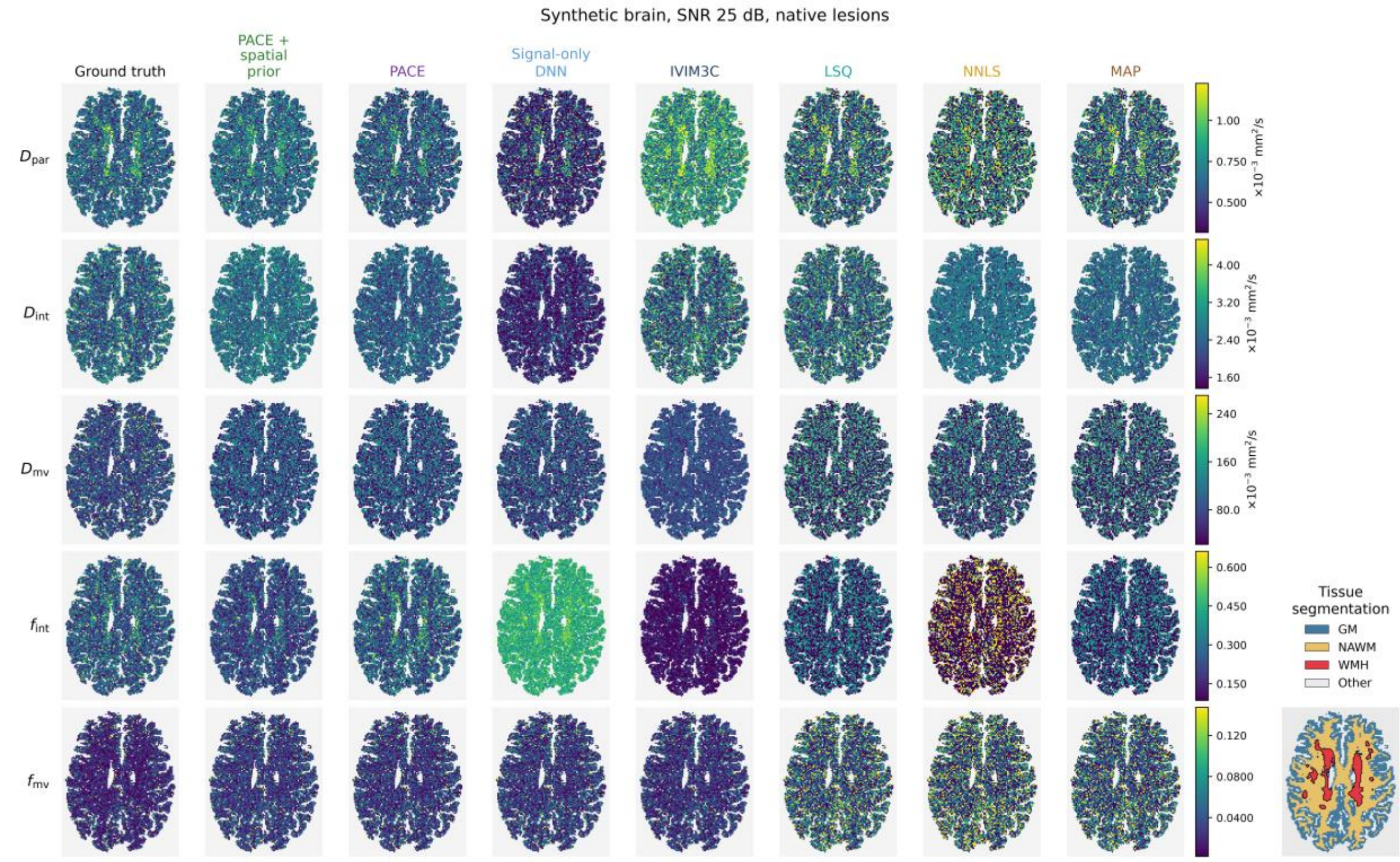

***Figure 8.*** *Predicted IVIM parameter maps for the hybrid brain at native WMH burden and SNR 25 dB. The hybrid brain combines in-vivo structural anatomy (T1w, FLAIR) with synthetic IVIM ground truth sampled from the training dictionary pool. Each column represents predictions from seven candidate models (PACE, PACE spatial prior, DNN, IVIM3C, LSQ, NNLS, MAP), while each row shows the corresponding IVIM parameter maps. Each parameter row's color scale reflects different prediction values, with higher color indicating higher parameter values, allowing direct visual comparison with the ground truth. The PACE variants showed greater agreement with ground truth across four out of five IVIM parameters, achieving lowest RMSE for* $D_{par}$, $D_{int}$, $f_{int}$, *and* $f_{mv}$. *IVIM-3C claimed best* $D_{mv}$ *reconstruction. The panel at the lower right shows the subject's tissue segmentation map. Abbreviation: PACE -* ***P****hysics-***A****nchored* ***C****ross-Attention* ***E****mbedding, DNN –* ***D****eep* ***N****eural* ***N****et, IVIM3C- Published IVIM deep net by Voorter et al.*[18]*, LSQ –* ***L****east* ***S****quare, NNLS –* ***N****on-***N****egative* ***L****east* ***S****quare,* ***MAP*** *–* ***M****aximum-***A****-***P****osteriori.*

**LIST OF SUPPLEMENTAL TABLES:**

**LIST OF SUPPLEMENTAL FIGURES:**

| **Method** | **Data set** | $D_{par}$ ($10^{-3}$ mm²/s) | $D_{int}$ ($10^{-3}$ mm²/s) | $D_{mv}$ ($10^{-3}$ mm²/s) | $f_{int}$ | $f_{mv}$ |
|---|---|---|---|---|---|---|
| PACE | Full | 0.193 | 0.695 | 64.9 | 0.134 | 0.0218 |
| | Degenerate | 0.190 | 0.722 | 61.9 | 0.149 | 0.0206 |
| PACE + spatial prior | Full | 0.175 | 0.669 | 60.9 | 0.126 | 0.0201 |
| | Degenerate | 0.172 | 0.695 | 57.0 | 0.141 | 0.0184 |
| Signal-only DNN | Full | 0.239 | 0.881 | 63.4 | 0.188 | 0.0221 |
| | Degenerate | 0.229 | 0.895 | 60.2 | 0.190 | 0.0210 |
| IVIM3C | Full | 0.252 | 0.861 | 56.1 | 0.210 | 0.0238 |
| | Degenerate | 0.259 | 0.882 | 51.0 | 0.231 | 0.0210 |
| LSQ | Full | 0.232 | 0.949 | 82.0 | 0.206 | 0.0552 |
| | Degenerate | 0.228 | 0.958 | 81.0 | 0.216 | 0.0500 |
| NNLS | Full | 0.404 | 0.858 | 77.2 | 0.301 | 0.0810 |
| | Degenerate | 0.387 | 0.836 | 75.9 | 0.312 | 0.0678 |
| MAP | Full | 0.222 | 0.827 | 80.4 | 0.206 | 0.0547 |
| | Degenerate | 0.219 | 0.825 | 79.4 | 0.215 | 0.0494 |

***Table S1:*** *Synthetic IVIM parameter RMSE performance for the full synthetic set and its degenerate subset data across SNR 20-35 dB. Cross-tissue degenerate pairs are defined as pairs with IVIM attenuation signals differ by less than 0.2% in signal root mean square (RMS) across b-values. PACE variants have the lowest parameter RMSEs, with the PACE retaining the lowest RMSE for four out five parameters except for* $D_{mv}$. *Abbreviation: PACE -* ***P****hysics-****A****nchored* ***C****ross-Attention* ***E****mbedding, DNN –* ***D****eep* ***N****eural* ***N****et, IVIM3C- Published IVIM deep net by Voorter et al.*[1]*, LSQ –* ***L****east* ***S****quare, NNLS –* ***N****on-****N****egative* ***L****east* ***S****quare,* ***MAP*** *–* ***M****aximum-****A****-****P****osteriori.*

| Method | Signal RMSE | Dpar RMSE ($10^{-3}$ mm²/s) | Dint RMSE ($10^{-3}$ mm²/s) | Dmv RMSE ($10^{-3}$ mm²/s) | Fint RMSE | Fmv RMSE |
|---|---|---|---|---|---|---|
| PACE + spatial prior | 0.0250 | 0.155 | 0.716 | 65.6 | 0.111 | 0.0235 |
| PACE | 0.0274 | 0.177 | 0.732 | 67.7 | 0.112 | 0.0240 |
| Signal-only DNN | 0.0278 | 0.237 | 0.978 | 67.4 | 0.207 | 0.0247 |
| IVIM3C | 0.0231 | 0.223 | 0.850 | 58.0 | 0.189 | 0.0240 |
| LSQ | 0.0253 | 0.233 | 0.957 | 86.0 | 0.204 | 0.0558 |
| NNLS | 0.0291 | 0.378 | 0.849 | 81.1 | 0.286 | 0.0731 |
| MAP | 0.0237 | 0.224 | 0.802 | 84.1 | 0.203 | 0.0553 |

**Table S2:** IVIM parameter RMSE performance for the hybrid brain data set at SNR 25 dB. PACE variants have the lowest parameter RMSEs, with the PACE retaining the lowest RMSE for four out five parameters except for $D_{mv}$. Interestingly, PACE with spatial prior only ranked second for signal RMSE, and IVIM3C ranked first despite having a higher param RMSE. This shows that signal RMSE and parameter RMSE do not always correlate. *Abbreviation: PACE - **P**hysics-**A**nchored **C**ross-Attention **E**mbedding, DNN – **D**eep **N**eural **N**et, IVIM3C- Published IVIM deep net by Voorter et al.[1], LSQ – **L**east **S**quare, NNLS – **N**on-**N**egative **L**east **S**quare, **MAP** – **M**aximum-**A**-**P**osteriori.*

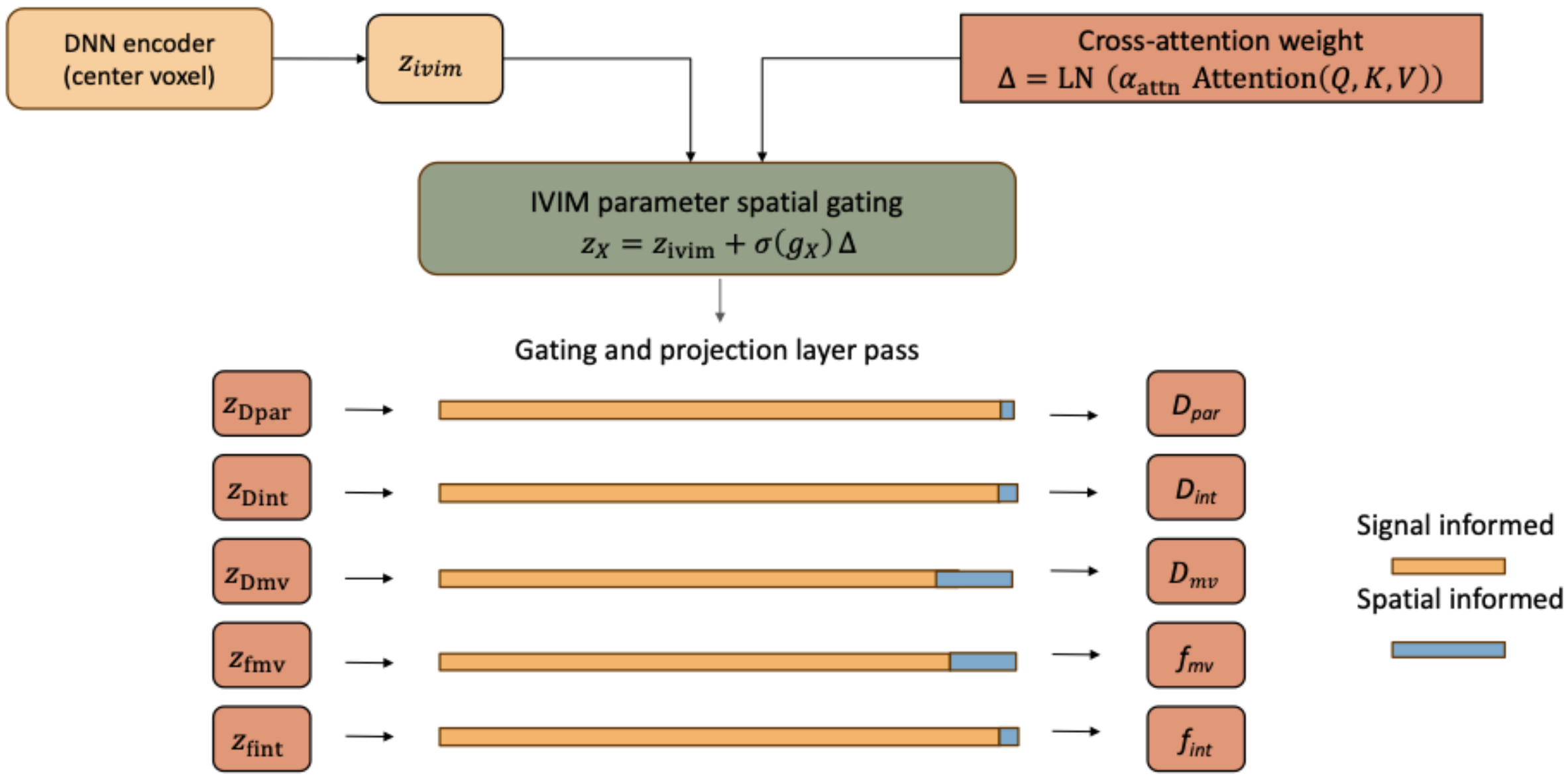

***Figure S1:*** *Spatial Gating Operation Details. IVIM embedding $z_{ivim}$ is combined with fused spatial representation from attention to produce parameter-specific representations $z_x$. During training, all gates are initialized to $\sigma(-2)$. Because the spatial gates are trained to minimize IVIM loss, spatial context is admitted only when it improves the signal fit, preventing the fusion output from overwhelming the physics-anchored signal. Bar proportions reflect gate values, where orange indicates the signal-driven estimation and blue indicates the spatial context (not to scale, for illustrative purpose only).*

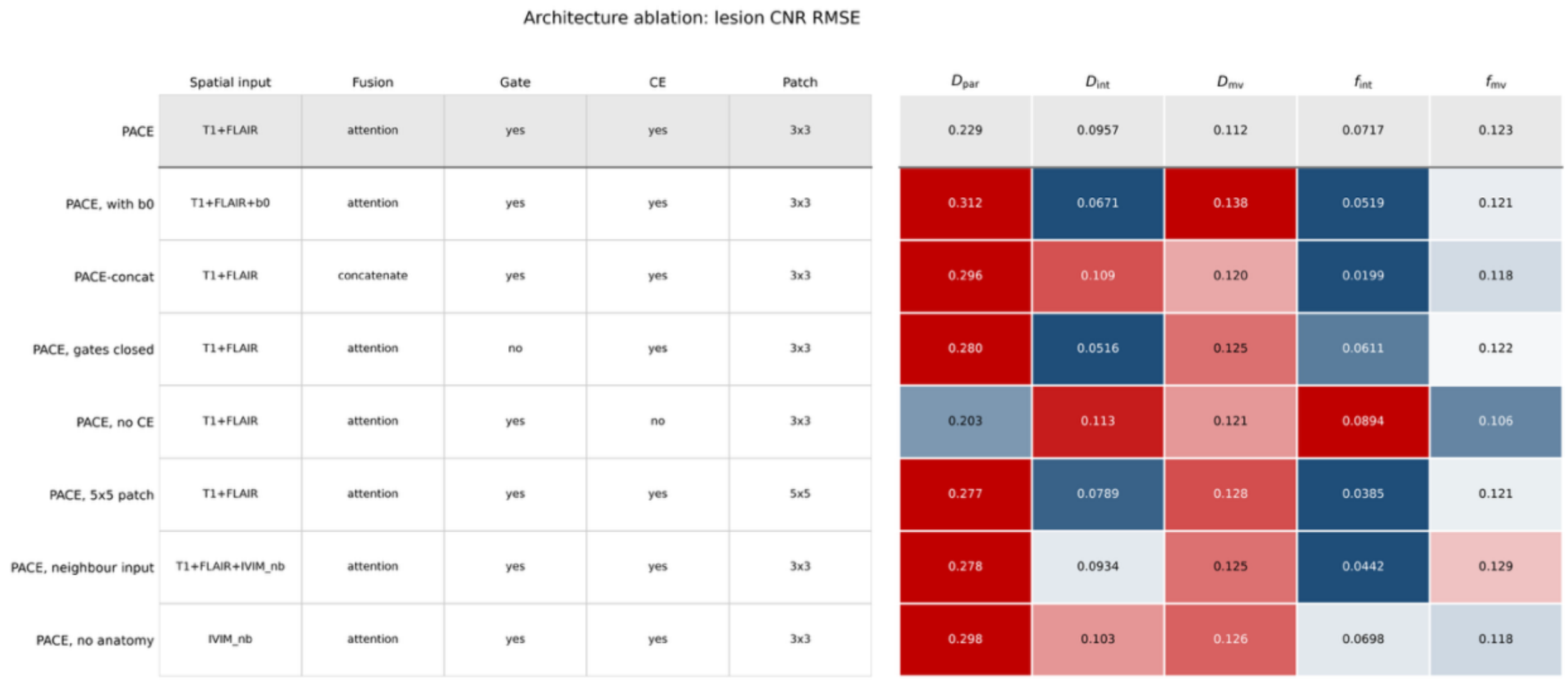

Architecture ablation: lesion CNR RMSE

| | Spatial input | Fusion | Gate | CE | Patch | $D_{par}$ | $D_{int}$ | $D_{mv}$ | $f_{int}$ | $f_{mv}$ |
|---|---|---|---|---|---|---|---|---|---|---|
| PACE | T1+FLAIR | attention | yes | yes | 3x3 | 0.229 | 0.0957 | 0.112 | 0.0717 | 0.123 |
| PACE, with b0 | T1+FLAIR+b0 | attention | yes | yes | 3x3 | 0.312 | 0.0671 | 0.138 | 0.0519 | 0.121 |
| PACE-concat | T1+FLAIR | concatenate | yes | yes | 3x3 | 0.296 | 0.109 | 0.120 | 0.0199 | 0.118 |
| PACE, gates closed | T1+FLAIR | attention | no | yes | 3x3 | 0.280 | 0.0516 | 0.125 | 0.0611 | 0.122 |
| PACE, no CE | T1+FLAIR | attention | yes | no | 3x3 | 0.203 | 0.113 | 0.121 | 0.0894 | 0.106 |
| PACE, 5x5 patch | T1+FLAIR | attention | yes | yes | 5x5 | 0.277 | 0.0789 | 0.128 | 0.0385 | 0.121 |
| PACE, neighbour input | T1+FLAIR+IVIM_nb | attention | yes | yes | 3x3 | 0.278 | 0.0934 | 0.125 | 0.0442 | 0.129 |
| PACE, no anatomy | IVIM_nb | attention | yes | yes | 3x3 | 0.298 | 0.103 | 0.126 | 0.0698 | 0.118 |

***Figure S2:*** *Lesion contrast fidelity as represented by CNR RMSE heatmap between ablation models. Across the architectural ablations, lesion CNR RMSE showed a consistent trade-off between $D_{par}$ /$D_{mv}$ and $f_{int}$ contrast accuracy, suggesting that the current spatial-context architecture could not simultaneously preserve diffusivity and fractional contrasts.*

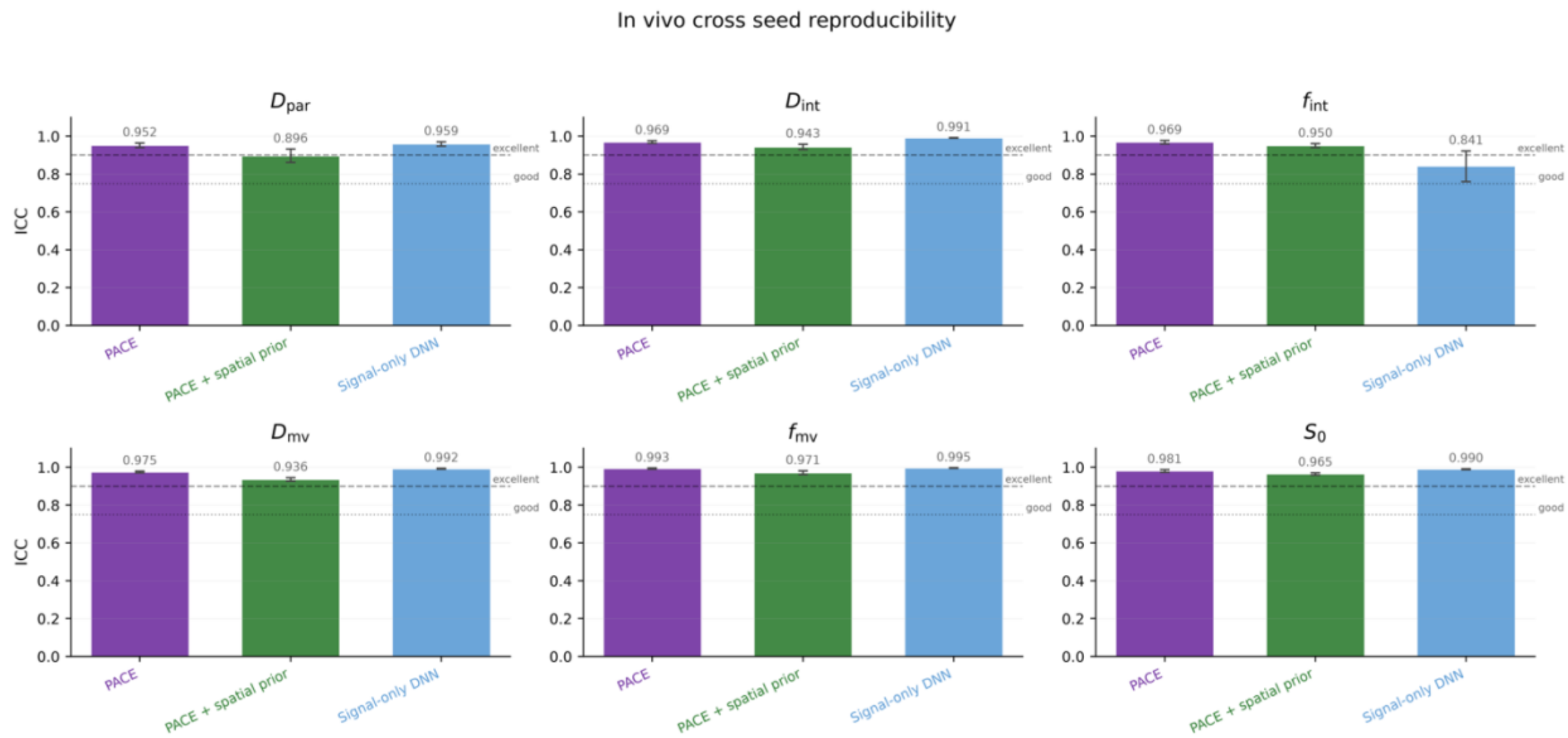

***Figure S3:*** *ICC comparison between default PACE (purple), PACE with spatial priors (green), and DNN (blue) for cross-seed reproducibility. All three models retained good-to-excellent ICC scores, with PACE models slightly trailing DNN for most IVIM parameters due to variants introduced through the spatial inputs. Abbreviation notes: PACE -* ***P****hysics-***A****nchored* ***C****ross-Attention* ***E****mbedding, DNN –* ***D****eep* ***N****eural* ***N****et.*

## SUPPLEMENTAL REFERENCES

1. Voorter PHM, Backes WH, Gurney-Champion OJ, Wong S-M, Staals J, van Oostenbrugge RJ, van der Thiel MM, Jansen JFA, Drenthen GS. Improving microstructural integrity, interstitial fluid, and blood microcirculation images from multi-b-value diffusion MRI using physics-informed neural networks in cerebrovascular disease. Magnetic Resonance in Med. 2023;90(4):1657–71. doi: 10.1002/mrm.29753.